\documentclass[journal]{IEEEtran}
\usepackage{cite}

\usepackage{graphicx}
\graphicspath{{/Figure/}}
\DeclareGraphicsExtensions{.eps,.pdf,.png,.jpg}

\usepackage{setspace}
\usepackage{array}

\usepackage{multirow,subfigure}
\usepackage{color}
\usepackage{longtable,balance}
\begin{document}

\title{Integrated Resource Management in Software Defined Networking, Caching and Computing}
\linespread{1.1}
\author{\IEEEauthorblockN{Qingxia~Chen\IEEEauthorrefmark{1}\IEEEauthorrefmark{2},
F.~Richard~Yu\IEEEauthorrefmark{3}, Tao~Huang\IEEEauthorrefmark{1}\IEEEauthorrefmark{2}, Renchao~Xie\IEEEauthorrefmark{1}\IEEEauthorrefmark{2}, Jiang~Liu\IEEEauthorrefmark{1}\IEEEauthorrefmark{2}, and Yunjie~Liu\IEEEauthorrefmark{1}\IEEEauthorrefmark{2}}\\
\IEEEauthorblockA{
\IEEEauthorrefmark{1}State Key Laboratory of Networking and Switching Technology, Beijing University of Post and Telecommunications, Beijing 100876, China \\
\IEEEauthorrefmark{2}Beijing Advanced Innovation Center for Future Internet Technology, Beijing 100876, China\\
\IEEEauthorrefmark{3}Depart. of Systems and Computer Eng., Carleton University, Ottawa, ON, Canada
}
}

\maketitle
\begin{abstract}
Recently, there are significant advances in the areas of networking, caching and computing. Nevertheless, these three important areas have traditionally been addressed separately in the existing research. In this paper, we present a novel framework that integrates networking, caching and computing in a systematic way and enables dynamic orchestration of these three resources to improve the end-to-end system performance and meet the requirements of different applications. Then, we consider the bandwidth, caching and computing resource allocation issue and formulate it as a joint caching/computing strategy and servers selection problem to minimize the combination cost of network usage and energy consumption in the framework.
To minimize the combination cost of network usage and energy consumption in the framework, we formulate it as a joint caching/computing strategy and servers selection problem.
In addition, we solve the joint caching/computing strategy and servers selection problem using an exhaustive-search algorithm. Simulation results show that our proposed framework significantly outperforms the traditional network without in-network caching/computing in terms of network usage and energy consumption.

\end{abstract}


\IEEEpeerreviewmaketitle
\section{Introduction}
Recently, there are significant advances in the areas of networking, caching and computing, which can have profound impacts on our society though the developments of smart cities, smart transportation, smart homes, etc. {Software-defined networking} (SDN) has been considered as one of the most promising technologies on realization of programmable networking, and has been deployed well in existing IP networks, such as Internet service providers and data center networks \cite{sdnopenissue,sdnsurvey}. By separating the control plane (decision functions) from the data plane (forwarding functions) in networks, SDN enables the development of new routing and forwarding approaches without the need to replace hardware components in the core network, simplifying network management and facilitating network evolution \cite{HHB14,YYG15,CYL15,CYY15,YY15}.

In the area of caching, {information-centric networking} (ICN), which has been extensively studied in recent years \cite{XVSFT01,LMT14}, enables {in-network caching} to reduce the duplicate content transmission in networks \cite{ZLZ15}. By focusing on the data's names instead of their locations, ICN provides native support for secure communication, scalable and efficient content distribution, and the enhanced capability for mobility \cite{FYH15,LY15ICC,LYY16,LYZ15,WYL16}. In the area of computing, {cloud computing} paradigm has been widely adopted to utilize the computing resources in remote provider's servers via the Internet, providing enterprises and end users with a range of application services and freeing them from the specification of many details \cite{AFG10,YYB15,CYB15,CYB14,CYB14_Infocom,YYB14,YL15}. Nevertheless, it is not feasible or economical to satisfy current applications requirements of mobility support, location awareness, low latency and big data analytics as the distance between the cloud and the edge device is usually large. To address these issues, {fog computing} has been proposed to extends cloud computing and services close to end devices \cite{Zha15,SCM15}. A similar technique, called {mobile edge computing}, is being standardized to allocate computing resources in wireless access networks \cite{MEC}.

\begin{figure*}[tb]
\centering
\includegraphics[width=16cm]{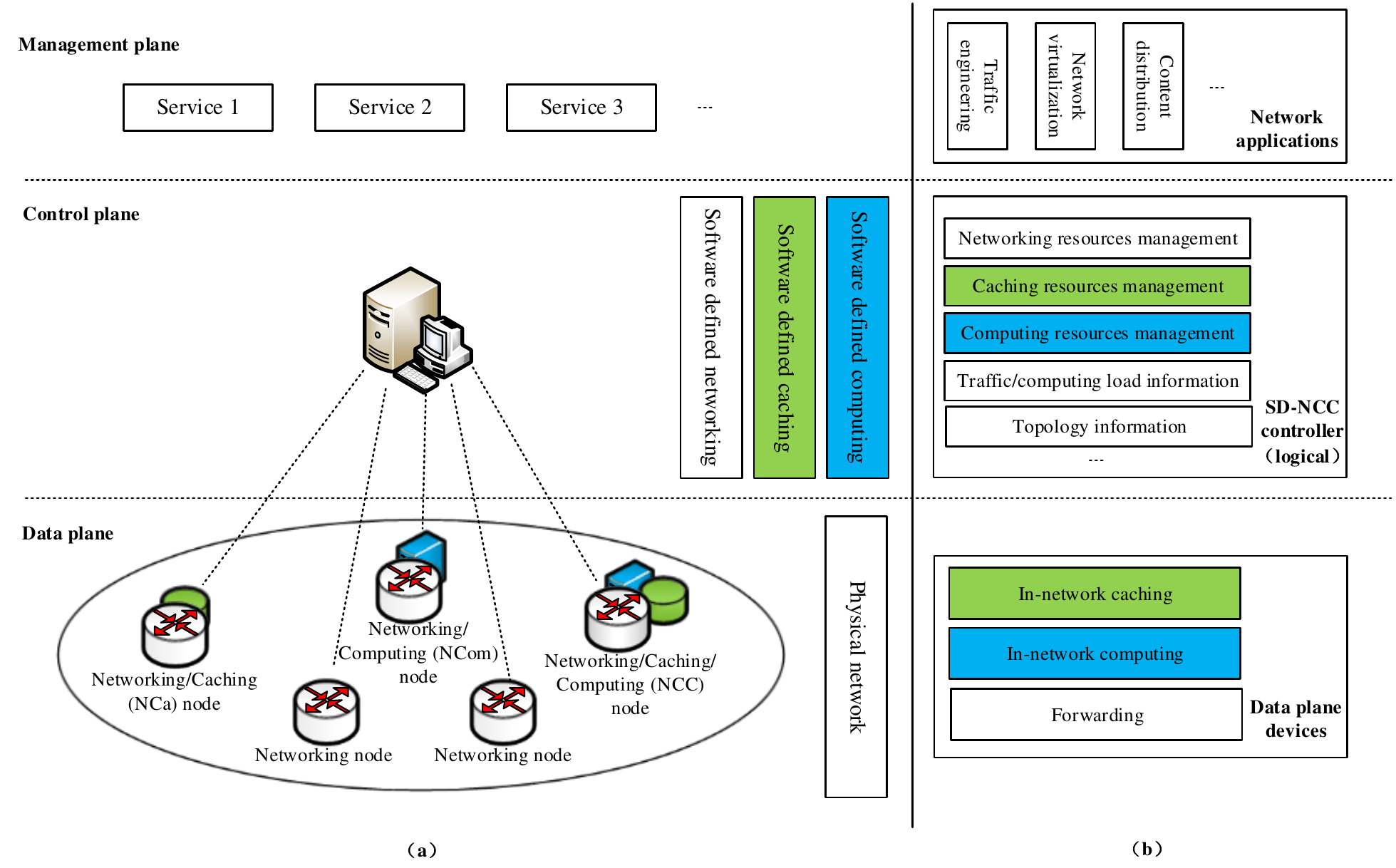}
\caption{Software-defined networking, caching, and computing in (a) planes and (b) system design architecture.}
\label{system architecture}
\end{figure*}

How to manage, control and optimize network, cache and compute (the three important underlying resources) can have great impacts on the performance of system and applications. Currently, in the works of SDN and ICN, these three resources are traditionally addressed separately, which could result in suboptimal performance. We present a novel framework called SD-NCC (Software Defined Networking, Caching and Computing) that integrates networking, caching and computing in a systematic way to meet the requirements of different applications and improve the end-to-end system performance.
As Fig.~\ref{system architecture} shows, the SD-NCC framework can be divided in three planes of functionality: data, control, and management planes (similar to SDN). Different from SDN, the added in-network caching and computing in the data plane become inherent fundamental capabilities in SD-NCC. Thus, besides the software defined networking, SD-NCC enables the software defined caching and computing. Additionally, each data packet carries a name and a signature (similar to ICN), thus enabling information centric paradigm.

Based on the SD-NCC framework, we present a comprehensive system model and formulate a joint caching/computing strategy and servers selection problem as a mixed-integer nonlinear programming (MINLP) problem to minimize the combination cost of network usage and energy consumption. Specifically, we formulate the caching/computing capacity allocation problem and derive the optimal deployment numbers of service copies. In addition, we develop an exhaustive-search algorithm to find the optimal caching/computing strategy and servers selection strategy. Simulation results show that compared with traditional network, our proposed SD-NCC framework significantly reduces the traffic traversing the network and has performance advantage on energy consumption cost. We also show the impact of service popularity on optimal deployment number of service copies.

The rest of the article is organized as follows. Section II presents a system model for this framework. Section III formulates the joint bandwidth, caching and computing resource allocation problem and work out a near-optimal caching/computing strategy and servers selection solution. Simulation results are discussed in Section IV. Finally, we conclude this paper in Section V.

\section{System Model\label{formulation}}
In this section, we present the system model for the SD-NCC framework.

\subsection{Network Model}
Excellent works have been done on modeling networks \cite{MYL04, YL01, LYH10, YK07, XYJL12, ATV12, YWL06_MONET,LY15, WYS10, GYJ10, BYC12, XYJ12, YTH09, YHT10, LYJ10, YKL06,YZX11, LYJ15,GYJ11,BY14,LYL09,YYG15,BYY15,ZYN12_JSAC,ZYL10,WTY14,BYL11_Online,BY13}. In this paper, Consider a network represented by graph $\mathbf{G} = \left\{ {\mathbf{V},\mathbf{E}} \right\}$, where $\mathbf{V}$ denotes the set of nodes and $\mathbf{E}$ denotes the set of directed physical links. A node can be a router, a user, or a server. We introduce the following notation:
\begin{itemize}
  \item $\mathbf{S} = \left\{ {{\mathbf{S}_A},\,{\mathbf{S}_B}} \right\}$, the set of in-network server nodes; ${\mathbf{S}_A} \subset \mathbf{V}$, the set of caching nodes; ${\mathbf{S}_B} \subset \mathbf{V}$, the set of computing nodes.
  \item $o$, a single virtual origin node which is always able to serve the service requests.
  \item $\mathbf{U}$, the set of users who issue service requests, $\mathbf{U} \subset \mathbf{V}$.
  \item $\mathbf{K} = \left\{ {{\mathbf{K}_A},\,{\mathbf{K}_B}} \right\}$, the set of service elements; ${\mathbf{K}_A}$, the set of content elements; ${\mathbf{K}_B}$, the set of computation services.
  \item $ \mathbf{M}:\,{\left\{ {m_u^k} \right\}_{(k,u) \in \mathbf{K} \times \mathbf{U}}}$, user demand matrix for duration $t$; ${m_u^k}$, the demand of  service $k$ from user $u$, $u \in \mathbf{U}$.
  \item (${\lambda _{k_1}^A}$, ${o_{k_1}^A}$), content $k_1$'s attribute, $k_1 \in {\mathbf{K}_A}$; ${\lambda _{k_1}^A}$, the number of content $k_1$'s requests for duration $t$; ${o_{k_1}^A}$, the size of content $k_1$.
  \item (${\lambda _{k_2}^A}$, ${o_{k_2}^B}$, ${c_{k_2}^B}$ ), computation $k_2$'s attribute, $k_2 \in {\mathbf{K}_B}$; ${\lambda _{k_2}^A}$, the number of computation $k_2$'s requests for duration $t$; ${o_{k_2}^B}$, the amount of computation $k_2$'s data communications, $k_2 \in {\mathbf{K}_B}$; ${c_{k_2}^B}$, the amount of computation $k_2$'s workload (which can be measured by the number of clock cycles or execution time).
\end{itemize}

\subsection{Caching/Computing Model}\label{cc model}
Let $c_i^A$, $c_i^B$ denote the caching capacity and computing capacity of server node $i$.
We assume the finite capacity of in-network server nodes and the infinite capacity of original server.
We use $\mathbf{H}:\left\{ {h_i^k} \right\}$ to denote the caching/computing strategy matrix, where $h_i^k = 1$ if server $i$ is able to provide the requested service $k$ and $h_i^k = 0$ if it is not. Specifically, $h_o^k = 1$.

Thus, the caching/computing capacity constraints are as follows:

\begin{equation}
\sum\nolimits_k {h_i^ko_k^A}  \le c_i^A\;,\forall i \in {\mathbf{S}_A}
\end{equation}
and
\begin{equation}
\sum\nolimits_k {h_i^kc_k^B}  \le c_i^B\;,\forall i \in {\mathbf{S}_B}
\end{equation}

\subsection{Servers Selection Model}\label{ss model}
Let $\mathbf{P}:\left\{ {\rho _{i,u}^k} \right\}$ denote the servers selection matrix, where $\rho _{i,u}^k \in \left[ {0,\;1} \right]$ denotes the proportion of the service request $k$ from user $u$ served by server $i$.
Node $u$ can be viewed as an edge router that aggregates the traffic of many endhosts, which may be served by different servers.
Let $ \mathbf{X}:\left\{ {x_{i,u}^k} \right\}$ denote the traffic allocation matrix, where ${x_{i,u}^k}$ denotes the traffic of service $k$ from user $u$ to server $i$ for duration $t$.
Combining with the caching/computing strategy, we have
\begin{equation}
x_{i,u}^k = m_u^k\rho _{i,u}^kh_i^k,\forall i \in S \cup \left\{ o \right\}
\end{equation}
and
\begin{equation}
  \sum\nolimits_i {\rho _{i,u}^kh_i^k}  = 1
\end{equation}

\subsection{Routing Model}
Let $\mathbf{R}:\left\{ {r_l^{i,u}} \right\}$ be the routing matrix, where $r_l^{i,u} \in \left[ {0,1} \right]$ denotes the proportion of traffic from user $u$ to server $i$ that traverses link $l$. $r_l^{i,u}$ equals 1 if the link $l$ is on the path from server $i$ to user $u$ and 0 otherwise.

Then the routing selection strategy is given by
\begin{equation}\label{equat: 1}
\sum\limits_{l:l \in In\left( v \right)} {r_l^{i,u}}  - \sum\limits_{l:l \in Out(v)} {r_l^{i,u}}  = {I_{v = i}},\;\;\forall i \in \mathbf{S} \cup \left\{ o \right\},v \in \mathbf{V}\backslash \mathbf{U}
\end{equation}
where ${I_{v = i}}$ is an indicator function which equals 1 if $v = i$ and 0 otherwise, ${In\left( v \right)}$ denotes the set of incoming links to node $v$, and ${Out\left( v \right)}$ denotes the set of outgoing links from node $v$.

The capacity of a link $l$ is $c_l$. Thus, the total traffic traversing link $l$ denoted by ${x_l}$ is given by
\begin{equation}\label{equat: 2}
{x_l} = \sum\nolimits_{k,i,u} {x_{i,u}^kr_l^{i,u}}  \le {c_l},\forall i \in \mathbf{S} \cup \left\{ o \right\}
\end{equation}

\subsection{Energy Model}\label{energy model}
The escalation of energy consumption in  networks directly results in the increase of greenhouse gas emission, which has been recognized as a major threat for environmental protection and sustainable development \cite{BY13,CYY12,ZYL10,BYL11_Online,BYC12,BY14}. The energy is mainly consumed by content caching, data computing and traffic transmission in SD-NCC framework. We discuss these three energy models as follows.

\subsubsection{Caching Energy}
$E_{ca}^A$, we use an energy-proportional model, similar to~\cite{CGK12}. If $n_{k_1}$ copies of the content $k_1$ are cached for duration $t$, the energy consumed by caching ${n_k}{o_k^A}$ bits is given by
\begin{equation}
E_{ca,k_1}^A = {n_{k_1}}{o_{k_1}^A}p_{ca}^At
\end{equation}
where $p_{ca}^A$ is the power efficiency of caching. The value of $p_{ca}^A$ strongly depends on the caching hardward technology.

\subsubsection{Computing Energy}
We assume that computation applications are executed within the virtual machines (VMs) which are deployed on the computing nodes and the amount of incoming computation ${k_2}$'s workloads from all users is ${\lambda _{k_2}^B}{c_{k_2}^B}$ for duration $t$.
The energy consumption of the computing nodes consists of two parts, including a dynamic energy consumption part $E_{active}^B$ when the VMs are active (processing computing service requests) and a static energy consumption part $E_{static}^B$ when the VMs are idle.

Thus, the dynamic energy consumption on processing the workloads is
\begin{equation}
{E_{active}^{B,{k_2}}} = {\lambda _{k_2}^B}{c_{k_2}^B}p_{active}^B
\end{equation}
and the static energy consumption for $m_{k_2}$ copies of computation ${k_2}$'s dedicated VM is
\begin{equation}
E_{static}^{B,{k_2}} = m_{k_2}p_{static}^Bt
\end{equation}
where $p_{active}^B$, $p_{static}^B$ are the average power efficiency of VMs in active and static states, respectively.

Note that the dynamic energy consumption is independent of the VMs' copies.

\subsubsection{Transmission Energy}

The transmission energy $E_{tr}$ consumption mainly consists of the energy consumption at routers and energy consumption along the links.

The transmission energy of content $E_{tr}^A$ and computation $E_{tr}^B$ for duration $t$ are given by
\begin{equation}
 E_{tr}^{\rm{A,{k_1}}} = {\lambda _{k_1}^A}{o_{k_1}^A}\left[ {{p_{tr,link}} \cdot {d_{k_1}^A} + {p_{tr,node}} \cdot \left( {{d_{k_1}^A} + 1} \right)} \right]
\end{equation}
\begin{equation}
  E_{tr}^{\rm{B,{k_2}}} = {\lambda _{k_2}^B}{o_{k_2}^B}\left[ {{p_{tr,link}} \cdot {d_{k_2}^B} + {p_{tr,node}} \cdot \left( {{d_{k_2}^B} + 1} \right)} \right]
\end{equation}
where ${p_{tr,link}}$ and ${p_{tr,node}}$ are the energy efficiency parameters of the link and node respectively. $d_{k_1}^A$ and $d_{k_2}^B$ represent the average hop distance to the content server nodes for content $k_1$'s request and the computation server nodes for computation $k_2$'s request, respectively.

\section{Caching/Computing/Bandwidth Resource Allocation}\label{decomposition}

In this subsection, we consider the joint caching, computing and bandwidth resource allocation problem and formulate it as a joint caching/computing strategy and servers selection problem (CCS-SS) which we show is a mixed integer nonlinear programming (MINLP) problem.
By relaxing the whole caching, computing, link constraints, we focus on the caching/computing capacity allocation for each service element and formulate it as a a nonlinear programming (NLP) problem. We derive the optimal caching/computing capacity allocation (i.e., the copies number for each service element) and then propose an exhaustive-search algorithm to find the optimal caching/computing strategy and servers selection.

\subsection{Problem Formulation}\label{wholeview}

\subsubsection{Objective Function}
Based on the known topology, traffic matrix, routing matrix and the energy parameters of network equipments, we introduce the following quantities:

\begin{itemize}
  \item $d_{i,u}$, the hop distance between server node $i$ and user $u$;
  \item $D_{i,u}$, the end-to-end latency between server node $i$ and user $u$;
  \item ${a_{i,u}} = {p_{tr,link}}{d_{i,u}} + {p_{tr,node}}\left( {{d_{i,u}} + 1} \right)$, the energy for transporting per bit from server node $i$ to user $u$;
  \item ${f_{e,tr}} = \sum\limits_{k \in \mathbf{K}} {\sum\limits_{i \in \mathbf{S} \cup \left\{ o \right\}} {\sum\limits_{u \in \mathbf{U}} {{a_{i,u}}m_u^k\rho _{i,u}^kh_i^k} } } $, the transmission energy for duration $t$;
  \item ${f_{e,ca}} = \sum\limits_{{k_1} \in {\mathbf{K}_A}} {\sum\limits_{i \in \mathbf{S}_A} {h_i^{{k_1}}o_{{k_1}}^A{p_{ca}^At}} } $, the caching energy for duration $t$;
  \item ${f_{e,com}} = \sum\limits_{{k_2} \in {\mathbf{K}_B}} {\sum\limits_{i \in \mathbf{S}_B} {h_i^{{k_2}}p_{static}^Bt} } + \sum\limits_{{k_2} \in {\mathbf{K}_B}} {\lambda _{k_2}^B}{c_{k_2}^B}p_{active}^B$, the energy consumption on computing nodes for duration $t$;

  \item ${f_{tr}} = \sum\limits_{k \in \mathbf{K}} {\sum\limits_{i \in \mathbf{S} \cup \left\{ o \right\}} {\sum\limits_{u \in \mathbf{U}} {{D_{i,u}}m_u^k\rho _{i,u}^kh_i^k} } } $, the network traffic for duration $t$.
\end{itemize}

We select a combining objective which balances the energy costs and the network usage costs for duration $t$. The cost function is given by
\begin{equation}
\begin{array}{l}
f = {f_{e,ca}}\left( {h_i^k} \right) + {f_{e,com}}\left( {h_i^k} \right) + {f_{e,tr}}\left( {\rho _{i,u}^kh_i^k} \right)\\
\quad \quad  + \gamma {f_{tr}}\left( \rho _{i,u}^kh_i^k \right)
\end{array}
\end{equation}
where $\gamma $ is the weight between the two costs. The first three parts constitute the energy consumption for duration $t$ and the last one is network traffic which denotes the network usage in this paper.

\subsubsection{Formulation}
In the following, we formulate the CCS-SS problem to
minimize the combination cost function. The optimization problem is shown as follows:
\begin{spacing}{1.2}
\begin{equation}\label{formulation1}
\begin{array}{l}
\min \;f\left( {h_i^k,\rho _{i,u}^k} \right)\\
s.t.\;C{\kern 1pt} 1:\quad \sum\nolimits_i {\rho _{i,u}^kh_i^k}  = 1\\
\;\;\;\;\;\;C2:\quad \sum\nolimits_k {h_i^ko_k^A}  \le c_i^A\;,\forall i \in {\mathbf{S}_A}\\
\;\;\;\;\;\;C3:\quad \sum\nolimits_k {h_i^kc_k^B}  \le c_i^B\;,\forall i \in {\mathbf{S}_B}\\
\;\;\;\;\;\;C4:\quad {x_l} = \sum\nolimits_{k,i,u} {m_u^k\rho _{i,u}^kh_i^kr_l^{i,u}}  \le {c_l}\;,\\
\;\;\;\;\;\;\;\;\;\;\;\;\;\;\;\;\;\;\forall i \in \mathbf{S} \cup \left\{ o \right\}\\
\;\;\;\;\;\;C5:\quad h_i^k \in \left\{ {0,\;1} \right\},\rho _{i,u}^k \in \left[ {0,\;1} \right]\;
\end{array}
\end{equation}
\end{spacing}

Constraints (1) specifies that user $u$'s demand rate for service $k$ can be served by several servers simultaneously depending on caching/computing strategy and servers selection strategy.
Constraints (2) and (3) specify the caching/computing resource capacity limits on in-network server nodes.
Data rates are subject to link capacity constraint (4).

Problem (\ref{formulation1}) is difficult to solve based on the following observations:
\begin{itemize}
  \item Both the objective function and feasible set of (\ref{formulation1}) are not convex due to the binary variables $h_i^k$ and the product relationship between $h_i^k$ and $\rho _{i,u}^k$.
  \item The size of the problem is very large. For instance, if there are $F$ service elements and $N$ network nodes, the number of variables $h_i^k$ is $N^{F}$. In future networks, the number of the service elements and network nodes will rise significantly.
\end{itemize}

As is well known, a MINLP problem is expected to be NP-hard \cite{BL12}, and a variety of algorithms are capable of solving instances with hundreds or even thousands of variables. However, it is very challenging to find the global optimum resource allocation in our problem, since the number of variables and the constraints grow exponentially. How to efficiently solve it is left for future research.
Instead, in the next subsection, we propose an NLP formulation from a different view which can be solved analytically.

\subsection{Caching/Computing Capacity Allocation}\label{deployment numbers}

In this section, we focus on the optimal caching/computing capacity allocated for each service $k$, but not the optimal cache location.
We denote $n_{k_1}$ as the number of content $k_1$'s copies and $m_{k_2}$ as the number of computation ${k_2}$'s dedicated VM copies. For simplicity, we remove the integer constraint of $n_{k_1}$ and $m_{k_2}$.
In a network with $N$ nodes, if service $k$ can be provided by $n$ server nodes, $N-n$ nodes have to access the service via one or more hops in a steady state.
For the irregular and asymmetric network, the authors in \cite{CGK12} take a semi-analytical approach in deriving the average hop distance to the servers,$d$, as a power-law function of $n$:
\begin{equation}\label{hopdistance}
 d\left( n \right) = A{\left( {{N \mathord{\left/
 {\vphantom {N n}} \right.
 \kern-\nulldelimiterspace} n}} \right)^\alpha }
\end{equation}

Thus, we have
\begin{equation}
d_{{{\rm{k}}_1}}^A = A{\left( {{N \mathord{\left/
 {\vphantom {N {{n_{{k_1}}}}}} \right.
 \kern-\nulldelimiterspace} {{n_{{k_1}}}}}} \right)^\alpha },d_{{k_2}}^B = A{\left( {{N \mathord{\left/
 {\vphantom {N {{m_{{k_2}}}}}} \right.
 \kern-\nulldelimiterspace} {{m_{{k_2}}}}}} \right)^\alpha }
\end{equation}
where $d_{k_1}^A$ and $d_{k_2}^B$ are the average hop distance to content $k_1$'s copies and computation $k_2$'s dedicated VM copies, respectively.

We assume the average end-to-end latency is proportional to the average hop distance and the scaling factor is $\eta$. Then the network traffic for duration $t$ is
\begin{equation}
\begin{array}{l}
{T_{total}}=\sum\limits_{{k_1} \in {K_A}} {T_{k_1}^{A}}+\sum\limits_{{k_2} \in {K_B}} {T_{k_2}^{B}} \\ = \eta \left( {\sum\limits_{{k_1} \in {K_A}} {\lambda _{{k_1}}^Ao_{{k_1}}^Ad_{{k_1}}^A}  + \sum\limits_{{k_2} \in {K_B}} {\lambda _{{k_2}}^Bo_{{k_2}}^Bd_{{k_2}}^B} } \right)
\end{array}
\end{equation}

According to energy model in Subsection~\ref{energy model}, the total energy consumption for duration $t$ is as follows:

\begin{equation}
  \begin{array}{l}
{E_{total}} = E_{total}^A + E_{total}^B\\
 = \sum\limits_{{k_1} \in {{\bf{K}}_A}} {\left( {E_{ca,{k_1}}^A\left( {{n_{{k_1}}}} \right) + E_{tr}^{A,{k_1}}\left( {{n_{{k_1}}}} \right)} \right)} \\
 + \sum\limits_{{k_2} \in {{\bf{K}}_B}} {\left( {E_{static}^{B,{k_1}}\left( {{m_{{k_2}}}} \right) + E_{active}^{B,{k_2}} + E_{tr}^{B,{k_2}}\left( {{m_{{k_2}}}} \right)} \right)}
\end{array}
\end{equation}

Thus, the caching/computing capacity allocation problem can be formulated as:

\begin{equation}
\begin{array}{*{20}{l}}
{\min \;{E_{total}} + \gamma {T_{total}}}\\
{s.t.\quad 1 \le {n_{{k_1}}} \le N}\\
{\quad \quad \;1 \le {m_{{k_2}}} \le N}
\end{array}
\end{equation}

We ignore the server capacity constraint by assuming sufficiently large server capacities, since we are more interested in the impact of caching/computing capacity allocation on total network cost
than the decision of load balancing among congested servers.

By using the Lagrangian dual method, the optimal $n_{k_1}$ and $m_{k_2}$, which are denoted as ${n_{k_1}^*}$ and ${m_{k_2}^*}$, can be derived as:
\begin{equation}
\begin{array}{l}
  n_{k_1}^* = \max [1,\min [n_{k_1}^o,N]]\;\\
  m_{k_2}^* = \max [1,\min [m_{k_2}^o,N]]
\end{array}
\end{equation}
where $n_{k_1}^o$ and $m_{k_2}^o$ are given by
\begin{equation}
n_{{k_1}}^o = {\left[ {\frac{{A\lambda _{{k_1}}^A{\alpha}\left( {{p_{tr,link}} + {p_{tr,node}} + \gamma \eta } \right)}}{{p_{ca}^At}}} \right]^{\frac{1}{{\alpha  + 1}}}}{N^{\frac{\alpha }{{\alpha  + 1}}}}
\end{equation}
\begin{equation}
m_{{k_2}}^o = {\left[ {\frac{{A\lambda _{{k_2}}^Ao_{{k_2}}^B{\alpha}\left( {{p_{tr,link}} + {p_{tr,node}} + \gamma \eta } \right)}}{{p_{static}^Bt}}} \right]^{\frac{1}{{\alpha  + 1}}}}{N^{\frac{\alpha }{{\alpha  + 1}}}}
\end{equation}

\subsection{The Exhaustive-search Algorithm}
The CCS-SS problem (\ref{formulation1}) can be denoted by minimizing $f\left( \mathbf{H},\mathbf{P}|\mathbf{M},\mathbf{R} \right)$ with the caching, computing, link constraints. Note that if the caching/computing strategy matrix $\mathbf{B}$ is pre-known, the formulation (\ref{formulation1}) will turn into minimizing $f\left(\mathbf{P}|\mathbf{M},\mathbf{R,\mathbf{H}} \right)$ with the link constraints which is a linear optimization problem.
In Subsection~\ref{deployment numbers}, we derived the optimal deployment copies ${n_{k_1}^*}$ for each content $k_1$, and ${m_{k_2}^*}$ for computation service $k_2$ without regard to copy locations in the network.
We assume that the element numbers of ${\mathbf{K}_A}$, ${\mathbf{K}_B}$ are $F_1$, $F_2$.
Thus, the number of all possible combination subsets of caching/computing strategy (copy locations) $Q$ are
\begin{equation}
 Q=\prod \limits_{{k_1} \in {{\bf{K}}_A}}{C_{N}^{n_{k_1}}} \prod \limits_{{k_2} \in {{\bf{K}}_B}}{C_{N}^{m_{k_2}}}
\end{equation}
which is significantly reduced in contrast with the original number ${2^{N\left( {{F_1}+{F_2}} \right)}}$.
In the following, we propose an exhaustive-search algorithm to find optimal resource allocation solutions for each service.

We denote the set of all possible combination subsets as
\begin{equation}
  \mathbf{\Phi}  = \left\{ {{\mathbf{H}_{1}},{\mathbf{H}_{2}}, \ldots ,{\mathbf{H}_{Q}}} \right\}
\end{equation}

Then, the resource allocation process is described as follows. After the controller selects a caching/computing strategy ${\mathbf{H}_q} \in \mathbf{\Phi}$, it minimizes $f\left(\mathbf{P}|\mathbf{M},\mathbf{R,\mathbf{H}} \right)$ with the link constraints: finding the optimal servers selection $\mathbf{P}$ for the service requests to minimize the combination cost function $f$.
By exhaustive searching, we choose the optimal caching/computing strategy $\mathbf{H}^*$ :
\begin{equation}\label{equat: optimalx}
{\mathbf{H}^*} = \arg \min \mathop f\limits_{{\mathbf{H}_q} \in \mathbf{\Phi} } \left( {\mathbf{P}\;\left| {\mathbf{M},\mathbf{R},{\mathbf{H}_q}} \right.} \right)
\end{equation}
and the corresponding $\mathbf{P}^*$ and $f^*$.

\section{Simulation Results and Discussions}
In this section, we conduct simulations based on a hypothetical United States backbone network US64 \cite{CGK12}, which is a representation of the topological characteristics of commercial networks. In the simulation, the parameters of the network are referred to \cite{BAH09}.
Energy density of a link is ${p_{tr,link}} = 0.15 \times {10^{ - 8}}J/bit$. Energy density of a router is in the order of ${p_{tr,node}} = 2 \times {10^{ - 8}}J/bit$. Power density of caching is $p_{ca}^A = 0.25 \times {10^{ - 8}}W/bit$. Power density of computation VM in the static state is $p_{static}^B = 50W$. We assume that both content and computation traffic demands are $1Gbps$.

\begin{figure}[t!]
\centering
\includegraphics[width=9cm]{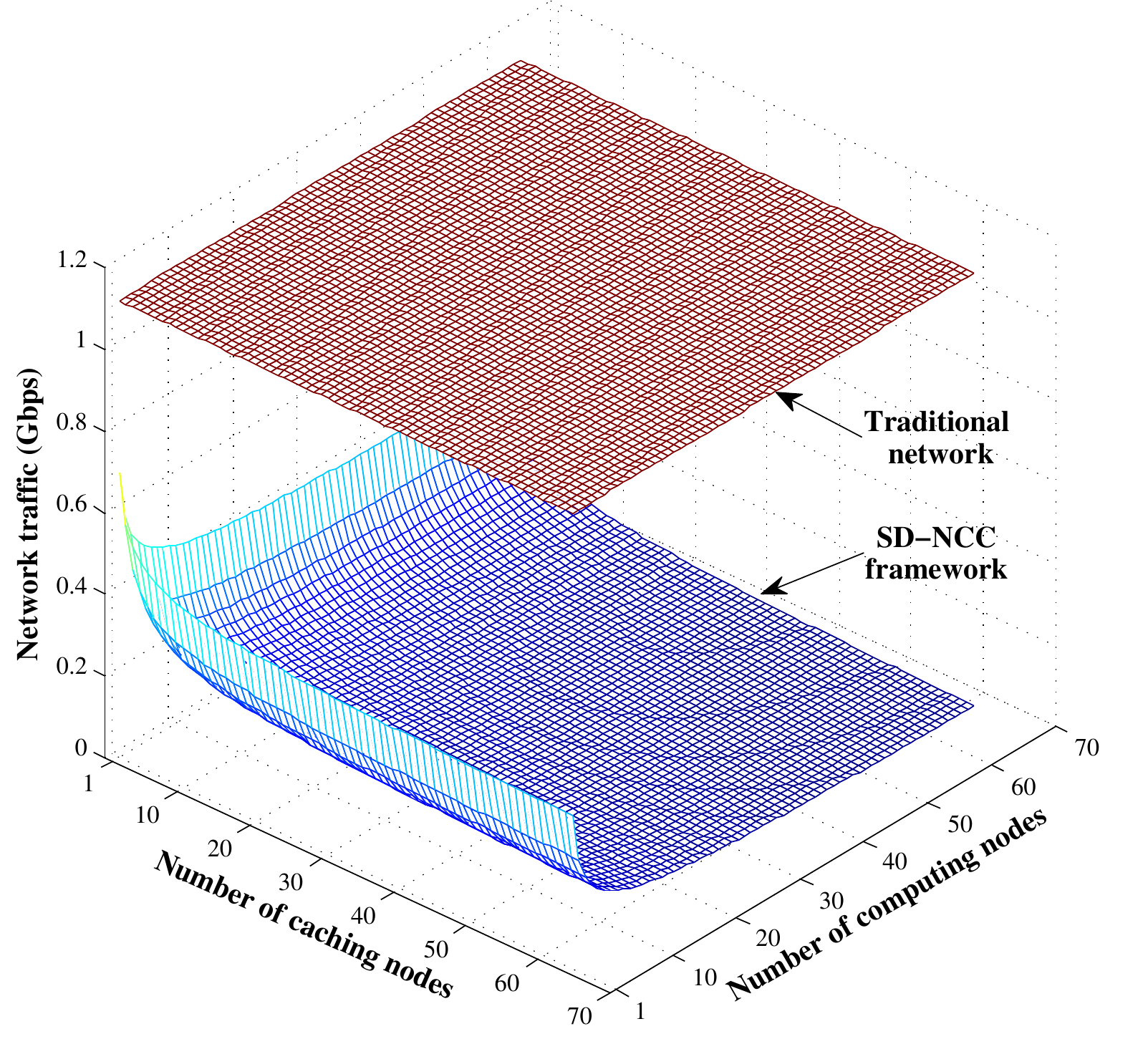}
\caption{The network traffic comparison between two schemes.}
\label{networktraffic}
\end{figure}


\begin{figure}[t!]
\centering
\includegraphics[width=8cm]{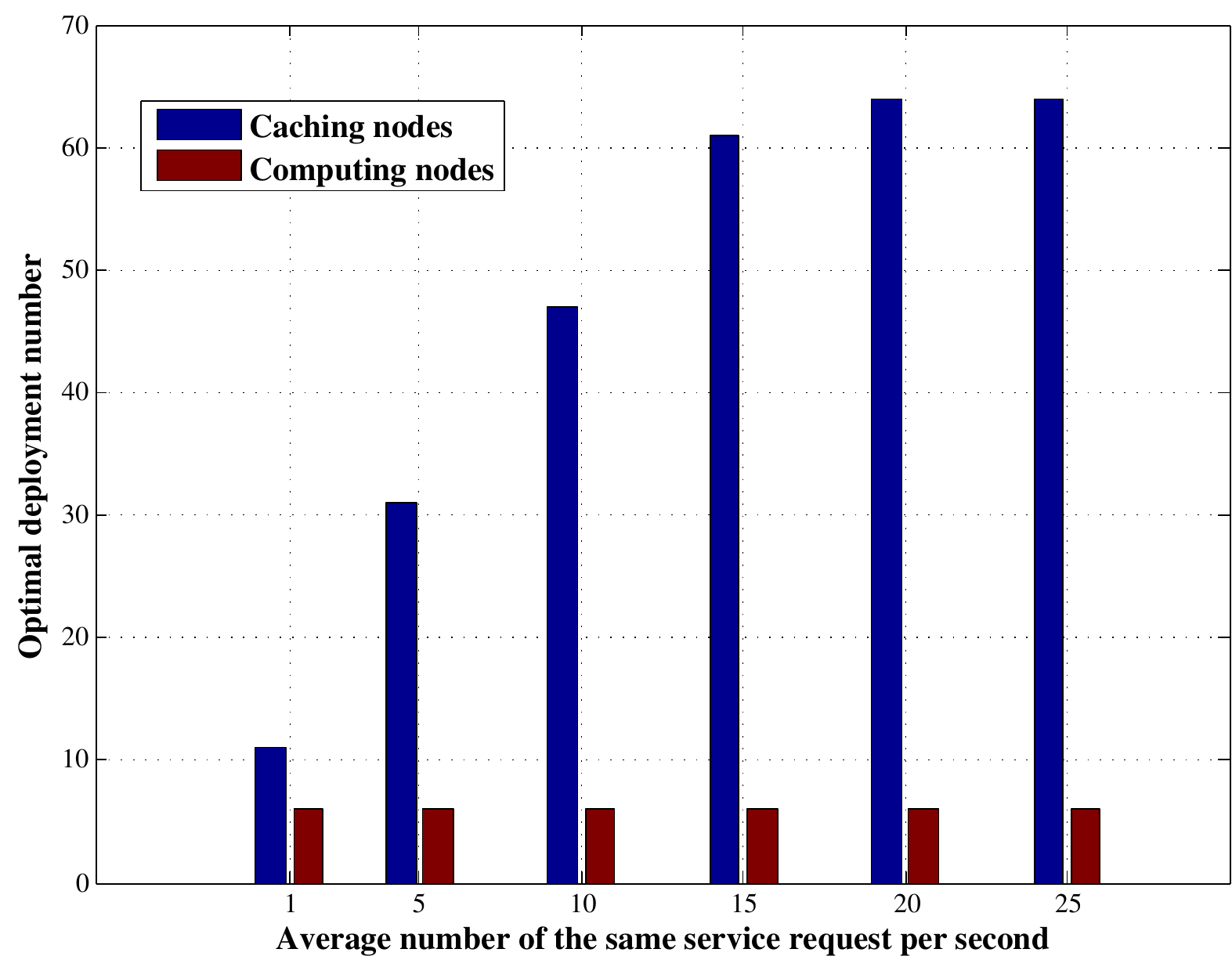}
\caption{The impact of service popularity on optimal deployment number of service copies.}
\label{optimaldeployment}
\end{figure}
\subsection{Network Usage Cost}
In Fig.~\ref{networktraffic}, we show the impact of caching/computing nodes' numbers on the average network traffic per second. We can see that SD-NCC architecture significantly reduces the network traffic compared with the traditional network without in-network caching/computing. This is due to the fact that with in-network caching and computing, large amount of content and computing requests are served on in-network nodes.


\subsection{Optimal Deployment Numbers}
In Subsection~\ref{deployment numbers}, we derive the optimal deployment numbers on minimizing the combination costs of energy consumption and network usage. Fig.~\ref{optimaldeployment} shows that under the fixed traffic demand, the optimal deployment number of caching copies for each content is proportional to the service popularity but the optimal deployment number of computation copies for each computation is independent of it. Because the larger number of the same content request (from different users), the less energy on caching (less content need to be cached). In contrast, the energy consumption on computing is fixed. Since computation tasks are more complex and the computation result is difficult to be used for other computation requests, even for the same computation requests from different users.

\section{Conclusions and Future Work}

In this paper, we  proposed to jointly consider networking, caching and computing in a systematic framework to improve the end-to-end system performance. We presented its system model and formulated the joint caching, computing and bandwidth resource allocation problem to minimize the energy cost and network usage cost. In addition, we derived the optimal deployment number of service copies and used an exhaustive-search algorithm to find the near-optimal caching/computing strategy and servers selection. Simulation showed that comparing the traditional network, our proposed SD-NCC framework significantly reduces the traffic traversing the network and has performance advantage on energy consumption. Future work is in progress to consider wireless networks in the proposed framework.

\section*{Acknowledgment}


 This work is jointly supported by the National High Technology Research and Development Program(863) of China (No. 2015AA016101), the National Natural Science Fund (No. 61302089, 61300184).

\ifCLASSOPTIONcaptionsoff
  \newpage
\fi

\balance
\bibliographystyle{IEEEtran}
\bibliography{ReferenceSCI,D:/CA/Papers/Ref}

\begin{thebibliography}{10}
\providecommand{\url}[1]{#1}
\csname url@samestyle\endcsname
\providecommand{\newblock}{\relax}
\providecommand{\bibinfo}[2]{#2}
\providecommand{\BIBentrySTDinterwordspacing}{\spaceskip=0pt\relax}
\providecommand{\BIBentryALTinterwordstretchfactor}{4}
\providecommand{\BIBentryALTinterwordspacing}{\spaceskip=\fontdimen2\font plus
\BIBentryALTinterwordstretchfactor\fontdimen3\font minus
  \fontdimen4\font\relax}
\providecommand{\BIBforeignlanguage}[2]{{%
\expandafter\ifx\csname l@#1\endcsname\relax
\typeout{** WARNING: IEEEtran.bst: No hyphenation pattern has been}%
\typeout{** loaded for the language `#1'. Using the pattern for}%
\typeout{** the default language instead.}%
\else
\language=\csname l@#1\endcsname
\fi
#2}}
\providecommand{\BIBdecl}{\relax}
\BIBdecl

\bibitem{sdnopenissue}
D.~Kreutz, F.~Ramos, P.~Esteves~Verissimo, C.~Esteve~Rothenberg,
  S.~Azodolmolky, and S.~Uhlig, ``Software-defined networking: A comprehensive
  survey,'' \emph{Proceedings of the IEEE}, vol. 103, no.~1, pp. 14--76, Jan.
  2015.

\bibitem{sdnsurvey}
W.~Xia, Y.~Wen, C.~H. Foh, D.~Niyato, and H.~Xie, ``A survey on
  software-defined networking,'' \emph{IEEE Commun. Surveys Tutorials},
  vol.~17, no.~1, pp. 27--51, Firstquarter 2015.

\bibitem{HHB14}
F.~Hu, Q.~Hao, and K.~Bao, ``A survey on software-defined network and openflow:
  From concept to implementation,'' \emph{IEEE Commun. Surveys Tutorials},
  vol.~16, no.~4, pp. 2181--2206, Fourthquarter 2014.

\bibitem{YYG15}
Q.~Yan, F.~R. Yu, Q.~Gong, and J.~Li, ``Software-defined networking ({SDN}) and
  distributed denial of service ({DDoS}) attacks in cloud computing
  environments: A survey, some research issues, and challenges,'' \emph{IEEE
  Commun. Survey and Tutorials}, vol.~18, no.~1, pp. 602--622, 2016.

\bibitem{CYL15}
Y.~Cai, F.~R. Yu, C.~Liang, B.~Sun, and Q.~Yan, ``Software defined
  device-to-device {(D2D)} communications in virtual wireless networks with
  imperfect network state information {(NSI)},'' \emph{IEEE Trans. Veh. Tech.},
  2015, dOI:10.1109/TVT.2015.2483558.

\bibitem{CYY15}
L.~Cui, F.~R. Yu, and Q.~Yan, ``When big data meets software-defined networking
  ({SDN}): {SDN} for big data and big data for {SDN},'' \emph{IEEE Network},
  vol.~30, no.~1, pp. 58--65, Jan. 2016.

\bibitem{YY15}
Q.~Yan and F.~R. Yu, ``Distributed denial of service attacks in
  software-defined networking with cloud computing,'' \emph{IEEE Commun. Mag.},
  vol.~53, no.~4, pp. 52--59, Apr. 2015.

\bibitem{XVSFT01}
G.~Xylomenos, C.~N. Ververidis, V.~A. Siris, N.~Fotiou, C.~Tsilopoulos,
  X.~Vasilakos, K.~V. Katsaros, and G.~C. Polyzos, ``A survey of
  information-centric networking research,'' \emph{IEEE Commun. Surveys
  Tutorials}, vol.~16, no.~2, pp. 1024--1049, Second Quarter 2014.

\bibitem{LMT14}
S.~Lederer, C.~Mueller, C.~Timmerer, and H.~Hellwagner, ``Adaptive multimedia
  streaming in information-centric networks,'' \emph{IEEE Network}, vol.~28,
  no.~6, pp. 91--96, Nov. 2014.

\bibitem{ZLZ15}
M.~Zhang, H.~Luo, and H.~Zhang, ``A survey of caching mechanisms in
  information-centric networking,'' \emph{IEEE Commun. Surveys Tutorials},
  vol.~17, no.~3, pp. 1473--1499, Thirdquarter 2015.

\bibitem{FYH15}
C.~Fang, F.~R. Yu, T.~Huang, J.~Liu, and Y.~Liu, ``A survey of green
  information-centric networking: Research issues and challenges,'' \emph{IEEE
  Commun. Surveys Tutorials}, vol.~17, no.~3, pp. 1455--1472, Thirdquarter
  2015.

\bibitem{LY15ICC}
C.~Liang and F.~R. Yu, ``Virtual resource allocation in information-centric
  wireless virtual networks,'' in \emph{Proc. IEEE ICC'15}, London, UK, June
  2015.

\bibitem{LYY16}
C.~Liang, F.~R. Yu, H.~Yao, and Z.~Han, ``Virtual resource allocation in
  information-centric wireless virtual networks,'' \emph{IEEE Trans.\ Veh.\
  Tech.}, 2016, to appear, available online.

\bibitem{LYZ15}
C.~Liang, F.~R. Yu, and X.~Zhang, ``Information-centric network function
  virtualization over {5G} mobile wireless networks,'' \emph{IEEE Network},
  vol.~29, no.~3, pp. 68--74, May 2015.

\bibitem{WYL16}
K.~Wang, F.~R. Yu, and H.~Li, ``Information-centric virtualized cellular
  networks with device-to-device ({D2D}) communications,'' \emph{IEEE Trans.
  Veh. Tech.}, 2016, accepted, online.

\bibitem{AFG10}
M.~Armbrust, A.~Fox, R.~Griffith, and A.~D. Joseph, ``A view of cloud
  computing,'' \emph{Commun. ACM}, vol.~53, no.~4, pp. 50--58, Apr. 2010.

\bibitem{YYB15}
Z.~Yin, F.~R. Yu, S.~Bu, and Z.~Han, ``Joint cloud and wireless networks
  operations in mobile cloud computing environments with telecom operator
  cloud,'' \emph{IEEE Trans. Wireless Commun.}, vol.~14, no.~7, pp. 4020--4033,
  July 2015.

\bibitem{CYB15}
Y.~Cai, F.~R. Yu, and S.~Bu, ``Dynamic operations of cloud radio access
  networks ({C-RAN}) for mobile cloud computing systems,'' \emph{IEEE Trans.
  Veh. Tech.}, 2015, doi: 10.1109/TVT.2015.2411739, online.

\bibitem{CYB14}
------, ``Cloud computing meets mobile wireless communications in next
  generation cellular networks,'' \emph{IEEE Network}, vol.~28, no.~6, pp.
  54--59, Nov. 2014.

\bibitem{CYB14_Infocom}
------, ``Joint dynamic cloud and wireless networks operations in the mobile
  cloud computing environment,'' in \emph{Proc. IEEE Infocom'14 Workshop on
  Mobile Cloud Computing}, Toronto, ON, Apr. 2014.

\bibitem{YYB14}
Z.~Yin, F.~R. Yu, and S.~Bu, ``Joint dynamic cloud and wireless networks
  operations in the mobile cloud computing environment,'' in \emph{Proc. IEEE
  Globecom'14}, Austin, TX, Dec. 2014.

\bibitem{YL15}
F.~R. Yu and V.~C.~M. Leung, \emph{Advances in Mobile Cloud Computing
  Systems}.\hskip 1em plus 0.5em minus 0.4em\relax New York: CRC Press, 2015.

\bibitem{Zha15}
M.~Zhanikeev, ``A cloud visitation platform to facilitate cloud federation and
  fog computing,'' \emph{Computer}, vol.~48, no.~5, pp. 80--83, May 2015.

\bibitem{SCM15}
S.~Sarkar, S.~Chatterjee, and S.~Misra, ``Assessment of the suitability of fog
  computing in the context of {Internet} of things,'' \emph{IEEE Trans. Cloud
  Computing}, vol.~PP, no.~99, 2015, online.

\bibitem{MEC}
{ETSI}, ``Mobile-edge computing – introductory technical white paper,''
  \emph{ETSI White Paper}, Sept. 2014.

\bibitem{MYL04}
L.~Ma, F.~Yu, V.~C.~M. Leung, and T.~Randhawa, ``A new method to support
  {UMTS/WLAN} vertical handover using {SCTP},'' \emph{IEEE Wireless Commun.},
  vol.~11, no.~4, pp. 44--51, Aug. 2004.

\bibitem{YL01}
F.~Yu and V.~C.~M. Leung, ``Mobility-based predictive call admission control
  and bandwidth reservation in wireless cellular networks,'' in \emph{Proc.
  IEEE INFOCOM'01}, Anchorage, AK, Apr. 2001.

\bibitem{LYH10}
Z.~Li, F.~R. Yu, and M.~Huang, ``A distributed consensus-based cooperative
  spectrum sensing in cognitive radios,'' \emph{IEEE Trans. Veh. Tech.},
  vol.~59, no.~1, pp. 383--393, Jan. 2010.

\bibitem{YK07}
F.~Yu and V.~Krishnamurthy, ``Optimal joint session admission control in
  integrated {WLAN} and {CDMA} cellular networks with vertical handoff,''
  \emph{IEEE Trans. Mobile Computing}, vol.~6, no.~1, pp. 126--139, Jan. 2007.

\bibitem{XYJL12}
R.~Xie, F.~R. Yu, H.~Ji, and Y.~Li, ``Energy-efficient resource allocation for
  heterogeneous cognitive radio networks with femtocells,'' \emph{IEEE Trans.
  Wireless Commun.}, vol.~11, no.~11, pp. 3910 --3920, Nov. 2012.

\bibitem{ATV12}
A.~Attar, H.~Tang, A.~Vasilakos, F.~R. Yu, and V.~Leung, ``A survey of security
  challenges in cognitive radio networks: Solutions and future research
  directions,'' \emph{Proceedings of the IEEE}, vol. 100, no.~12, pp.
  3172--3186, 2012.

\bibitem{YWL06_MONET}
Y.~Fei, V.~W.~S. Wong, and V.~C.~M. Leung, ``Efficient {QoS} provisioning for
  adaptive multimedia in mobile communication networks by reinforcement
  learning,'' \emph{Mob. Netw. Appl.}, vol.~11, no.~1, pp. 101--110, Feb. 2006.

\bibitem{LY15}
C.~Liang and F.~R. Yu, ``Wireless network virtualization: A survey, some
  research issues and challenges,'' \emph{IEEE Commun. Surveys Tutorials},
  vol.~17, no.~1, pp. 358--380, Firstquarter 2015.

\bibitem{WYS10}
Y.~Wei, F.~R. Yu, and M.~Song, ``Distributed optimal relay selection in
  wireless cooperative networks with finite-state {Markov} channels,''
  \emph{IEEE Trans. Veh. Tech.}, vol.~59, no.~5, pp. 2149 --2158, June 2010.

\bibitem{GYJ10}
Q.~Guan, F.~R. Yu, S.~Jiang, and G.~Wei, ``Prediction-based topology control
  and routing in cognitive radio mobile ad hoc networks,'' \emph{IEEE Trans.
  Veh. Tech.}, vol.~59, no.~9, pp. 4443 --4452, Nov. 2010.

\bibitem{BYC12}
S.~Bu, F.~R. Yu, Y.~Cai, and P.~Liu, ``When the smart grid meets
  energy-efficient communications: Green wireless cellular networks powered by
  the smart grid,'' \emph{IEEE Trans.\ Wireless Commun.}, vol.~11, pp.
  3014--3024, Aug. 2012.

\bibitem{XYJ12}
R.~Xie, F.~R. Yu, and H.~Ji, ``Dynamic resource allocation for heterogeneous
  services in cognitive radio networks with imperfect channel sensing,''
  \emph{IEEE Trans. Veh. Tech.}, vol.~61, pp. 770--780, Feb. 2012.

\bibitem{YTH09}
F.~R. Yu, H.~Tang, M.~Huang, Z.~Li, and P.~C. Mason, ``Defense against spectrum
  sensing data falsification attacks in mobile ad hoc networks with cognitive
  radios,'' in \emph{Proc. IEEE Military Commun. Conf. (MILCOM)'09}, Oct. 2009.

\bibitem{YHT10}
F.~R. Yu, M.~Huang, and H.~Tang, ``Biologically inspired consensus-based
  spectrum sensing in mobile ad hoc networks with cognitive radios,''
  \emph{IEEE Network}, vol.~24, no.~3, pp. 26 --30, May 2010.

\bibitem{LYJ10}
C.~Luo, F.~R. Yu, H.~Ji, and V.~C.~M. Leung, ``Cross-layer design for {TCP}
  performance improvement in cognitive radio networks,'' \emph{IEEE Trans. Veh.
  Tech.}, vol.~59, no.~5, pp. 2485--2495, 2010.

\bibitem{YKL06}
F.~Yu, V.~Krishnamurthy, and V.~C.~M. Leung, ``Cross-layer optimal connection
  admission control for variable bit rate multimedia traffic in packet wireless
  {CDMA} networks,'' \emph{IEEE Trans.\ Signal Proc.}, vol.~54, no.~2, pp.
  542--555, Feb. 2006.

\bibitem{YZX11}
F.~R. Yu, P.~Zhang, W.~Xiao, and P.~Choudhury, ``Communication systems for grid
  integration of renewable energy resources,'' \emph{IEEE Network}, vol.~25,
  no.~5, pp. 22 --29, Sept. 2011.

\bibitem{LYJ15}
G.~Liu, F.~R. Yu, H.~Ji, V.~C.~M. Leung, and X.~Li, ``In-band full-duplex
  relaying for {5G} cellular networks with wireless virtualization,''
  \emph{IEEE Network}, vol.~29, no.~6, pp. 54--61, Nov. 2015.

\bibitem{GYJ11}
Q.~Guan, F.~R. Yu, S.~Jiang, and V.~Leung, ``Capacity-optimized topology
  control for {MANETs} with cooperative communications,'' \emph{IEEE Trans.\
  Wireless Commun.}, vol.~10, no.~7, pp. 2162 --2170, July 2011.

\bibitem{BY14}
S.~Bu and F.~R. Yu, ``Green cognitive mobile networks with small cells for
  multimedia communications in the smart grid environment,'' \emph{IEEE Trans.
  Veh. Tech.}, vol.~63, no.~5, pp. 2115--2126, June 2014.

\bibitem{LYL09}
J.~Liu, F.~R. Yu, C.-H. Lung, and H.~Tang, ``Optimal combined intrusion
  detection and biometric-based continuous authentication in high security
  mobile ad hoc networks,'' \emph{IEEE Trans. Wireless Commun.}, vol.~8, no.~2,
  pp. 806--815, 2009.

\bibitem{BYY15}
S.~Bu, F.~R. Yu, and H.~Yanikomeroglu, ``Interference-aware energy-efficient
  resource allocation for heterogeneous networks with incomplete channel state
  information,'' \emph{IEEE Trans. Veh. Tech.}, vol.~64, no.~3, pp. 1036--1050,
  Mar. 2015.

\bibitem{ZYN12_JSAC}
L.~Zhu, F.~R. Yu, B.~Ning, and T.~Tang, ``Cross-layer handoff design in
  {MIMO}-enabled {WLANs} for communication-based train control ({CBTC})
  systems,'' \emph{IEEE J. Sel. Areas Commun.}, vol.~30, no.~4, pp. 719--728,
  May 2012.

\bibitem{ZYL10}
S.~Zhang, F.~R. Yu, and V.~Leung, ``Joint connection admission control and
  routing in {IEEE} 802.16-based mesh networks,'' \emph{IEEE Trans.\ Wireless
  Commun.}, vol.~9, no.~4, pp. 1370 --1379, Apr. 2010.

\bibitem{WTY14}
Z.~Wei, H.~Tang, F.~R. Yu, M.~Wang, and P.~Mason, ``Security enhancements for
  mobile ad hoc networks with trust management using uncertain reasoning,''
  \emph{IEEE Trans. Veh. Tech.}, vol.~63, no.~9, pp. 4647--4658, Nov. 2014.

\bibitem{BYL11_Online}
S.~Bu, F.~R. Yu, and P.~Liu, ``Dynamic pricing for demand-side management in
  the smart grid,'' in \emph{Proc. IEEE Online Conference on Green
  Communications (GreenCom)'11}, Sept. 2011, pp. 47--51.

\bibitem{BY13}
S.~Bu and F.~R. Yu, ``A game-theoretical scheme in the smart grid with
  demand-side management: Towards a smart cyber-physical power
  infrastructure,'' \emph{IEEE Trans. Emerging Topics in Computing}, vol.~1,
  no.~1, pp. 22--32, June 2013.

\bibitem{CYY12}
G.~Cili, H.~Yanikomeroglu, and F.~R. Yu, ``Cell switch off technique combined
  with coordinated multi-point ({CoMP}) transmission for energy efficiency in
  beyond-{LTE} cellular networks,'' in \emph{Proc. IEEE ICC'12}, June 2012.

\bibitem{CGK12}
N.~Choi, K.~Guan, D.~C. Kilper, and G.~Atkinson, ``In-network caching effect on
  optimal energy consumption in content-centric networking,'' in \emph{Proc.
  IEEE ICC'12}, June 2012, pp. 2889--2894.

\bibitem{BL12}
S.~Burer and A.~N. Letchford, ``Non-convex mixed-integer nonlinear programming:
  a survey,'' \emph{Surveys in Operations Research and Management Science},
  vol.~17, no.~2, pp. 97--106, 2012.

\bibitem{BAH09}
J.~Baliga, R.~Ayre, K.~Hinton, and R.~S. Tucker, ``Architectures for
  energy-efficient iptv networks,'' in \emph{Optical Fiber Communication
  Conference (OFC)}, Mar. 2009, pp. 1--3.

\end{thebibliography}
\end{document}